\documentclass[aps,prd,preprint,superscriptaddress,tightenlines,nofootinbib]{revtex4}

\usepackage{graphicx}
\usepackage{dcolumn}
\usepackage{bm}
\usepackage{epsfig}
\newcommand{\vcs}{V_{cs}}
\newcommand{\lcsl}{\Lambda_{c}^{+} \rightarrow \Lambda e^+ \nu_e}

\begin{document}

\preprint{CLNS 04/1901}       
\preprint{CLEO 04-19}         

\title{Improved Measurement of the Form Factors in the Decay
    $\Lambda_{c}^{+} \rightarrow \Lambda e^+ \nu_e$}


\author{J.~W.~Hinson}
\author{G.~S.~Huang}
\author{J.~Lee}
\author{D.~H.~Miller}
\author{V.~Pavlunin}
\author{R.~Rangarajan}
\author{B.~Sanghi}
\author{E.~I.~Shibata}
\author{I.~P.~J.~Shipsey}
\affiliation{Purdue University, West Lafayette, Indiana 47907}
\author{D.~Cronin-Hennessy}
\author{C.~S.~Park}
\author{W.~Park}
\author{J.~B.~Thayer}
\author{E.~H.~Thorndike}
\affiliation{University of Rochester, Rochester, New York 14627}
% ...
\collaboration{CLEO Collaboration} 
\noaffiliation


\date{December 29, 2004}

\begin{abstract} 

Using the CLEO detector at the Cornell Electron Storage Ring, we
have studied the distribution of kinematic variables in the decay
$\lcsl$. By performing a four-dimensional maximum likelihood fit,
we determine the form factor ratio, $R = f_{2}/f_{1}= -0.31 \pm
0.05{\rm (stat)} \pm 0.04{\rm (syst)}$, the pole mass, $M_{\rm
pole}=(2.21 \pm 0.08{\rm (stat)} \pm 0.14{\rm (syst)})$~GeV/$c^2$,
and  the decay asymmetry parameter of the $\Lambda_c^+$,
$\alpha_{\Lambda_{c}} = -0.86 \pm {0.03}{\rm (stat)} \pm 0.02{\rm
(syst)}$, for $\langle q^2 \rangle = 0.67$~(GeV/$c^2$)$^2$. We
compare the angular distributions of the $\Lambda_c^+$ and
$\overline{\Lambda}_c^-$ and  find  no evidence for $CP$-violation:
$\mathcal{A}_{\Lambda_{c}} = \frac{(\alpha_{\Lambda_{c}} +
\alpha_{ \overline{\Lambda}_{c} } ) } {(\alpha_{\Lambda_{c}} -
\alpha_{ \overline{\Lambda}_{c} } ) } = 0.00 \pm 0.03{\rm (stat)}
\pm 0.01{\rm (syst)} \pm 0.02$, where the
third error is from  the uncertainty in the world average of the
$CP$-violating parameter, $\mathcal{A}_{\Lambda}$, for $\Lambda
\rightarrow p \pi^-$.

\end{abstract}

\maketitle


The charm quark is unstable and decays via a first order weak
interaction. In semileptonic decays, which are analogs of neutron
$\beta$ decay, the charm quark disintegrates predominantly into a strange
quark, a positron and a neutrino. 
The rate depends on the weak quark mixing
Cabibbo-Kobayashi-Maskawa (CKM) matrix element
$|V_{cs}|$ and strong interaction effects,
parameterized by form factors, which come into play because the
charm quark is bound with light quarks to form a meson or baryon.
Charm semileptonic decays allow a measurement of the form factors
because $|V_{cs}|$ is tightly constrained by
the unitarity of the CKM matrix~\cite{PDG2004}.

Within the framework of Heavy Quark Effective Theory
(HQET)~\cite{LambdaInHQET}, semileptonic ($J^P = 1/2^+ \rightarrow
1/2^+$) transitions of $\Lambda$-type baryons are simpler than
mesons as they consist of a heavy quark and a spin and isospin
zero light diquark. 
This simplicity leads to more reliable predictions~\cite{heavyToLight,KKModel} 
for form factors in heavy-to-light transitions.
The measurement of form factors in 
the $\Lambda_{c}^{+} \rightarrow \Lambda e^+ \nu_e$ transition provides 
a test  of HQET predictions in
the charm baryon sector, a test of Lattice QCD, and information
for the determination of the CKM matrix elements $|V_{cb}|$ and
$|V_{ub}|$ using $\Lambda_{b}^0$ decays
since HQET relates the form factors in $\Lambda_{c}^{+}$
semileptonic decays to those governing $\Lambda_{b}^0$
semileptonic transitions.

In the limit of negligible lepton mass, the semileptonic ($J^P = 1/2^+ \rightarrow 1/2^+$)
transition of a $\Lambda$-type baryon is parameterized in terms of four form factors:
two axial form factors, $F_{1}^{A}$ and $F_{2}^{A}$, and two
vector form factors, $F_{1}^{V}$ and $F_{2}^{V}$.
These form factors are functions of $q^2$, the invariant mass squared
of the virtual $W^+$.  The decay can be described in terms of helicity
amplitudes $H_{\lambda_{\Lambda}\lambda_{W}} = H_{\lambda_{\Lambda}\lambda_{W}}^{V}
+ H_{\lambda_{\Lambda}\lambda_{W}}^{A}$,
where $\lambda_{\Lambda}$ and $\lambda_{W}$ are the helicities of the
$\Lambda$ and $W^+$. The helicity amplitudes are related to the form
factors in the following way~\cite{KKModel}
\begin{eqnarray}
\sqrt{q^2}H^{V}_{\frac{1}{2}0} & = & \sqrt{Q_{-}}\;[(M_{\Lambda_c}+
M_{\Lambda})F^{V}_{1}-q^{2}F^{V}_{2}], \nonumber \\
H^{V}_{\frac{1}{2}1}&=&\sqrt{2Q_{-}}\;[-F^{V}_{1} + (M_{\Lambda_c}+
M_{\Lambda})F^{V}_{2}], \\
\sqrt{q^2}H^{A}_{\frac{1}{2}0} & = & \sqrt{Q_{+}}\;[(M_{\Lambda_c}-
M_{\Lambda})F^{A}_{1}+q^{2}F^{A}_{2}], \nonumber \\
H^{A}_{\frac{1}{2}1}&=&\sqrt{2Q_{+}}\;[-F^{A}_{1} - (M_{\Lambda_c}-
M_{\Lambda})F^{V}_{2}], \nonumber
\end{eqnarray}
\noindent where $Q_{\pm} = (M_{\Lambda_c} \pm M_{\Lambda})^{2} -q^2$.
The remaining helicity amplitudes can be obtained using the parity relations
$H^{V(A)}_{-\lambda_{\Lambda} -\lambda_{W}} = +(-) H^{V(A)}_{\lambda_{\Lambda}\lambda_{W}}$.
In terms of the helicity amplitudes, the decay angular distribution can be
written as~\cite{KKModel}~\cite{noteOnPlus}
\begin{widetext}
\begin{eqnarray}
\Gamma_S && = \frac{d{\Gamma}}{dq^{2}d\cos{\theta_{\Lambda}}d\cos{\theta_W}d\chi}  =
\mathcal{B}(\Lambda \rightarrow p \pi^-) \frac{1}{2} \frac{G^{2}_{F}}{(2\pi)^4}|V_{cs}|^{2}
\frac{q^{2}P}{24M^{2}_{\Lambda_c}} \times \nonumber \\  \label{rateTheory}
&& \{\frac{3}{8}(1 - \cos{\theta_W})^{2}|H_{\frac{1}{2}1}|^{2}(1+
\alpha_{\Lambda}\cos{\theta_{\Lambda}})+ \frac{3}{8}(1+\cos{\theta_W})^2|H_{-\frac{1}{2}-1}|^{2}(1-
\alpha_{\Lambda}\cos{\theta_{\Lambda}}) \\
&& + \frac{3}{4}\sin^{2}{\theta_W}[|H_{\frac{1}{2}0}|^{2}(1+\alpha_{\Lambda}\cos{\theta_{\Lambda}})+
|H_{-\frac{1}{2}0}|^{2}(1-\alpha_{\Lambda}\cos{\theta_{\Lambda}})] \nonumber \\
&& - \frac{3}{2\sqrt{2}} \alpha_{\Lambda}\cos{\chi}\sin{\theta_W}\sin{\theta_{\Lambda}}[(1 -
\cos{\theta_{W}})Re(H_{-\frac{1}{2}0}H_{\frac{1}{2}1}^{*})+ \nonumber
(1+ \cos{\theta_W})Re(H_{\frac{1}{2}0}H_{-\frac{1}{2}-1}^{*})]\}, \nonumber
\end{eqnarray}
\end{widetext}
\noindent where $G_F$ is the Fermi coupling constant, $|\vcs|$ is
a CKM matrix element, $P$ is the magnitude of the $\Lambda$ momentum 
in the $\Lambda^{+}_{c}$ rest frame, $\theta_{\Lambda}$ is the angle
between the momentum vector of the proton in the $\Lambda$  rest
frame and the $\Lambda$  momentum in the $\Lambda^{+}_{c}$ rest
frame, $\theta_W$ is the angle between the momentum vector of the
positron in the $W^+$ rest frame and the $\Lambda$ momentum in the
$\Lambda^{+}_{c}$ rest frame, $\chi$ is the angle between the
decay planes of the $\Lambda$ and $W^+$, and $\alpha_{\Lambda}$ is
the $\Lambda \rightarrow p \pi^-$ decay asymmetry parameter
measured to be $0.642  \pm 0.013$~\cite{PDG2004}.

In HQET, the heavy flavor and spin symmetries imply
relations among the form factors and reduce their number to one
when the decay involves only heavy quarks. For heavy-to-light transitions,
two form factors are needed to describe the hadronic current.
In this Letter, we follow Ref.~\cite{KKModel}, in which the $c$ quark
is treated as heavy and the $s$ quark as light. Two independent form factors $f_1$ and $f_2$
are related to the standard form factors in the following way
$F^{V}_{1}(q^2) = -F^{A}_{1}(q^2) = f_{1}(q^2)+\frac{M_{\Lambda}}{M_{\Lambda_c}}f_{2}(q^2)$
and
$F^{V}_{2}(q^2) = -F^{A}_{2}(q^2) = \frac{1}{M_{\Lambda_c}}f_{2}(q^2)$.
In general, $f_2$ is expected to be negative and smaller in magnitude than $f_1$.
If the $s$ quark is treated as heavy, $f_2$ is zero.

In order to extract the form factor ratio $R = f_{2}/f_{1}$ from a
fit to the decay rate, $\Gamma_S$, an assumption must be made about
the $q^2$ dependence of the form factors. The model of K\"{o}rner and
Kr\"{a}mer (KK)~\cite{KKModel} uses the  dipole form $f(q^2) =
\frac{f(q^2_{\rm max})}{(1-q^2/M^2_{\rm pole})^2}(1-\frac{q^2_{\rm
max}}{M^2_{\rm pole}})^2$ for both form factors, where the pole
mass is taken from the naive pole dominance model: $M_{\rm pole} =
m_{D^{*}_{s}} = 2.11$~GeV/$c^2$.

In this Letter, we perform, for the first time,  a simultaneous 
fit for the form factor ratio and pole mass in the decay $\lcsl$, 
and we make a first search for $CP$-violation in this decay.
The data sample used in this study was collected with the
CLEO~II~\cite{cleo2} and upgraded CLEO~II.V~\cite{cleo25}
detectors operating at the Cornell Electron Storage Ring (CESR).
The integrated  luminosity consists of 13.7~fb$^{-1}$ taken at and
just below the $\Upsilon(4S)$ resonance, corresponding to
approximately $18 \times 10^6$ $e^+ e^- \rightarrow c
\overline{c}$ events.    Throughout this paper charge conjugate
states are implicitly included, unless otherwise indicated, and
the symbol $e$ is used to denote an electron or positron.

The analysis is an extension of the technique described
in~\cite{ffstudy,lambdacStudy}. The decay $\Lambda_{c}^{+}
\rightarrow \Lambda e^+ \nu_e$ is reconstructed by detecting a
$\Lambda e^+$ pair with invariant mass in the range $m_{\Lambda
e^+} < m_{\Lambda_c}$. The positron is required to come from the
region of the event vertex. To reduce the background from $B$
decays, we require $R_2 = H_2 / H_0 > 0.2$, where $H_i$ are
Fox-Wolfram event shape variables~\cite{FWMoments}. Positrons are
identified using a likelihood function, which incorporates
information from the calorimeter and $dE/dx$ systems. The minimum
allowed momentum for positron candidates is 0.7~GeV/$c$, as the
positron fake rates are much higher at lower momentum. Positrons
must be detected in the region: $|\cos{\theta}| < 0.7$, where
$\theta$ is the angle between the positron momentum and the beam
line. The $\Lambda$ is reconstructed in the decay mode $\Lambda
\rightarrow p \pi^-$. The $\Lambda$ baryon is long-lived~($c\tau =
7.89$~cm), accordingly, the $\Lambda$ vertex is required  to be
greater than 5~mm from the primary vertex in the $r - \phi$ plane,
but the $\Lambda$ momentum must extrapolate to the primary vertex.
The $dE/dx$ measurement of the proton is required to be consistent
with the expected value. Combinations that satisfy interpretation
as a $K_{S}^{0}$ are rejected. The magnitude of the $\Lambda$
momentum is required to be greater than 0.8~GeV/$c$ in order to
reduce combinatorial background. These $\Lambda$ candidates are
then combined with Right Sign (RS) tracks consistent with
positrons, and the sum of the $\Lambda$ and $e^+$ momenta is
required to be greater than 1.4~GeV/$c$ in order to reduce the
background from $B$ decays.

The above selection criteria permit the isolation of signal events
with low background. The number of events passing the selection is
4060, of which $123 \pm 12$ are consistent with fake $\Lambda$
background,  $338 \pm 67$ with $\Xi_c \rightarrow \Xi e^+ \nu$
feed through and $398 \pm 58$ with $e$~fake  background. The
sidebands of the $p \pi^-$ invariant mass distribution are used to
estimate the fake $\Lambda$ background.  The background from $\Xi_c
\rightarrow \Xi e^+ \nu$ decays is estimated using the result of a
previous CLEO analysis~\cite{cascadeCStudy}.

The normalization and momentum spectrum of the $e$~fake background
is estimated using the Wrong Sign (WS) $h^+ \overline{\Lambda}$
data sample (no charge conjugation is implied), where $h^+$  and
$\overline{\Lambda}$ satisfy all analysis selection criteria. The
$h^+$ tracks in this sample are mostly fakes as there are few
processes contributing $e^+ \overline{\Lambda}$ pairs
after the selection criteria are applied. If no particle
identification is used for $h$, the probability to find a  
($h^+ \overline{\Lambda}$ or $h^- \Lambda$)~WS or ($h^+  \Lambda$ or 
$h^-  \overline{\Lambda}$)~RS pair is approximately equal
because the net charge of the event is zero. 
The equality is not perfect due to (a)  baryon conservation: a $\Lambda$ 
is more likely to be produced with an antiproton in WS rather than RS
combinations, and  (b) associated strangeness production: there 
is a higher fraction of kaons in RS rather than WS combinations. 
When the particle identification requirements for $h$ are applied, 
the importance of these correlations is magnified by the high $e$~fake 
rates of antiprotons and kaons. Therefore, antiprotons are
excluded by using only one WS charge conjugate state ($h^+
\overline{\Lambda}$), and the momentum region where the $e$~fake
rate from kaons is high is excluded by requiring $|\vec{ p_e }| >
0.7$~GeV/$c$. Differences that remain between the momentum spectra
and particle species of hadronic tracks $h^+ \overline{\Lambda}$
and $h^+ \Lambda$ are second order and are accounted for by a
systematic uncertainty.

Calculating kinematic variables requires knowledge of the
$\Lambda_{c}^+$  momentum, which is unknown due to the undetected neutrino.
The direction of the $\Lambda_{c}^+$ is approximated using the information
provided by the thrust axis of the event and the kinematic constraints of the decay.
The  magnitude of the $\Lambda_{c}^+$ momentum is obtained as a weighted average
of the roots of the quadratic equation $\vec{p}^{~2}_{\Lambda_c} =
(\vec{p}_{\Lambda} + \vec{p}_{e} + \vec{p}_{\nu})^2$.  The weights
are assigned based on the measured fragmentation function of $\Lambda_{c}^{+}$.
After the $\Lambda_{c}^{+}$ momentum is estimated, the four kinematic
variables are easily obtained. The kinematic variables  
$t \equiv q^2/q^{2}_{\rm max}$, $\cos{\theta_{\Lambda}}$, $\cos{\theta_{W}}$
and $\chi$ achieve resolutions of 0.2, 0.3, 0.2 and $45^\circ$, respectively.

A four-dimensional maximum likelihood fit in a manner similar
to Ref.~\cite{fitMethod} is performed in the space of
$t$, $\cos{\theta_{\Lambda}}$, $\cos{\theta_{W}}$
and $\chi$. The technique makes possible a multidimensional fit
to variables modified by experimental acceptance and resolution
taking into account correlations among the variables.
We have performed two types of fit. The first fit is unbinned in all
four dimensions. The second fit is unbinned in $\cos{\theta_{\Lambda}}$,
$\cos{\theta_{W}}$ and $\chi$, and binned in $t$.
While both fits produce consistent results,
the second fit is used for the main results.

The signal probability density function for the likelihood
function is estimated at each data point using signal Monte Carlo
(MC) events, generated according to the HQET consistent KK model
with a GEANT based simulation~\cite{geant}, by  sampling  the MC
distribution at the reconstructed level in a search volume around
the data point. The volume size is chosen so that the systematic
effect from finite search volumes is small and the required number
of MC events is not prohibitively high. The background probability
density functions are modeled similarly using samples of events
for each background component. The $e$~fake background is modeled
using a sample of events collected for that purpose from the data.
Feed through background from $\Xi_c \rightarrow \Xi e^+ \nu$ is
modeled by the MC sample generated according to the HQET
consistent KK model. Fake $\Lambda$ background is modeled using
the data events in the sidebands of the $p \pi^-$ invariant mass
distribution. For the binned part of the fit, the above
distributions are projected onto $t$ and binned. The background
normalizations are fixed in the fits to the measured values.

\begin{figure}[h]
\begin{center}
      \epsfig{figure=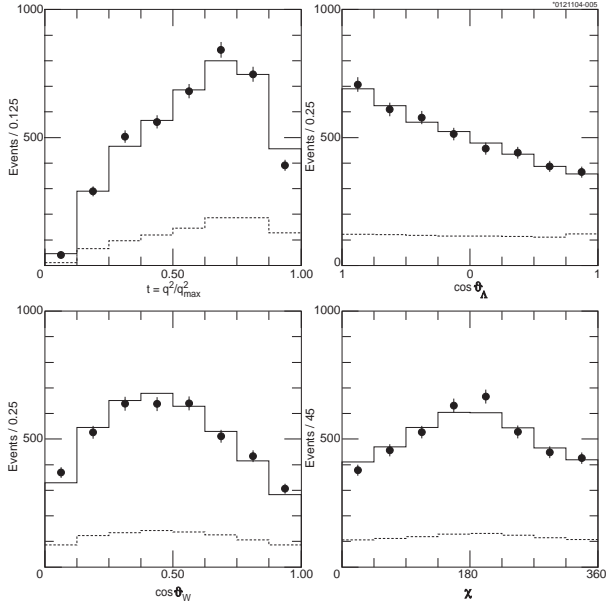,height=3.2in}
\end{center}
   \caption{ Projections of the data (points with statistical 
error bars) and the fit
(solid histogram) onto $t$, $\cos\theta_{\Lambda}$, $\cos\theta_W$ and $\chi$.
The dashed lines show the sum of the background distributions.
}
   \label{projections1}
\end{figure}

\begin{figure}
\begin{center}
      \epsfig{figure=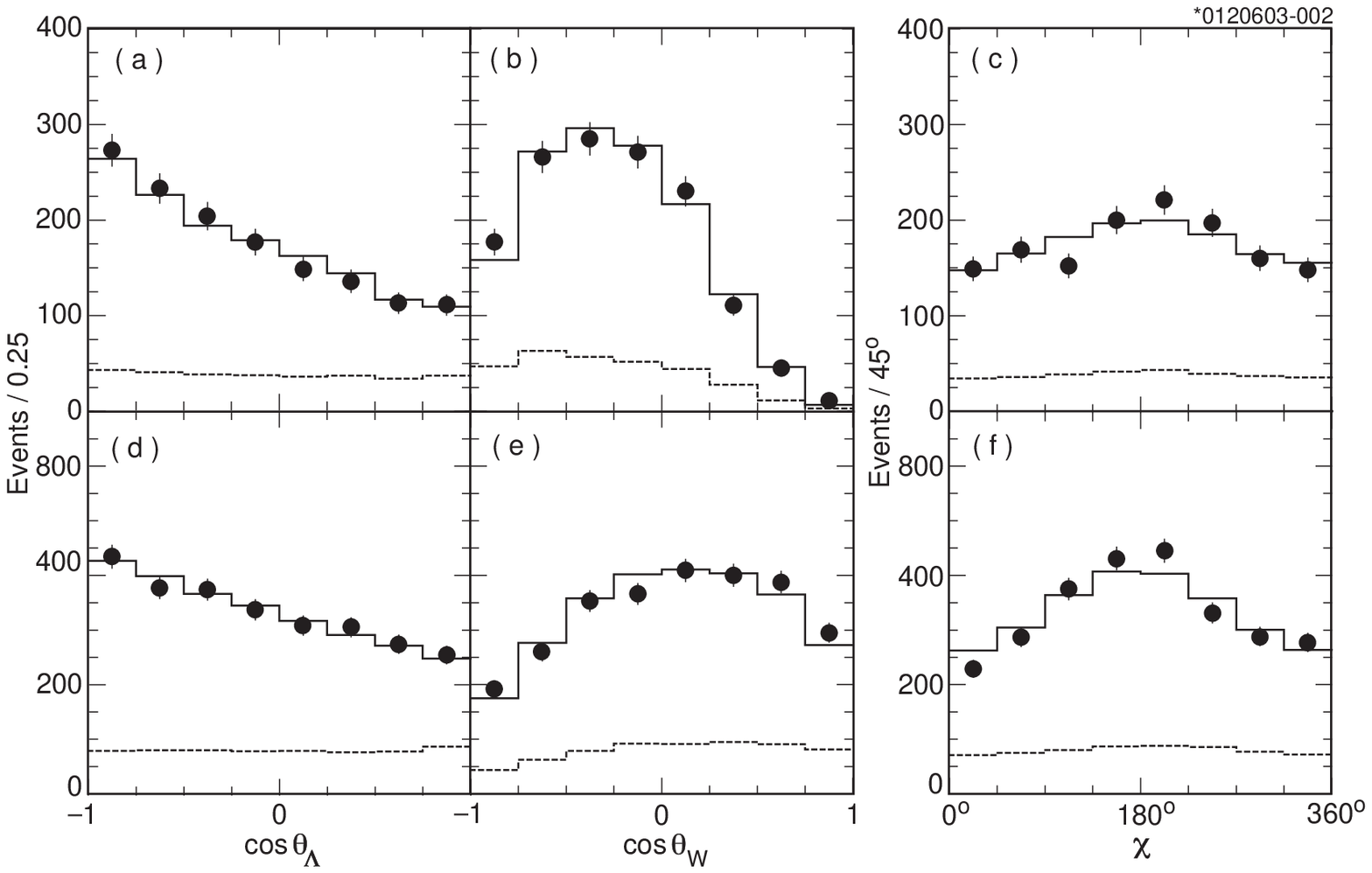,height=2.15in}
\end{center}
   \caption{ Projections of the data (points with statistical 
error bars) and the fit (solid histogram) onto $\cos\theta_{\Lambda}$, 
$\cos\theta_W$ and $\chi$  for two $t$ regions. The plots labeled (a), (b) 
and (c) are for $t < 0.5$; (d), (e) and (f) are for  $t > 0.5$. The dashed
lines show the sum of the background distributions.
}
\label{q2BinsProjections}
\end{figure}

Using the above method, a simultaneous fit for
the form factor ratio and the pole mass is made.
We find
$ 
R = -0.31 \pm 0.05 {\rm (stat)}
{\rm \;and\;}
M_{\rm pole} = (2.21 \pm 0.08 {\rm (stat)}
){\rm \;GeV/}c^2
$. 
This is the main result of the analysis.
Figures~\ref{projections1} and~\ref{q2BinsProjections} show the
$t$, $\cos\theta_{\Lambda}$, $\cos\theta_W$ and $\chi$ projections
for the data and fit.

We have considered the following sources of systematic uncertainty
and give our estimate of their magnitude in parentheses for $R$
and $M_{\rm pole}$, respectively. The uncertainty associated with
the size of the search volume is measured from a statistical
experiment in which a set of mock data samples, including signal
and all background components, was fit in the same way as the
data~(0.006, 0.048). The uncertainty due to the limited size of
the signal MC sample is estimated by dividing the sample into four
independent equal subsamples and repeating the fit (0.007, 0.012).
The uncertainty due to background normalizations is determined by
varying the estimated number of background events by one standard
deviation separately for each type of background~(0.023, 0.024).
The uncertainty associated with the modeling of the background
shapes,  including uncertainties originating from the modeling of
the $e$~fake background, and the unknown form factor ratio and
$M_{\rm pole}$ for the decays  $\Xi_c \rightarrow \Xi e^+ \nu$, is
estimated by varying these shapes or by using alternative
background samples~(0.024, 0.049). The uncertainty due to the
small background contribution of random $\Lambda e^+$ pairs from
the continuum ($e^+ e^- \rightarrow q \overline{q}$) and
$\Upsilon(4S) \rightarrow B \overline{B}$ events, which are not
modeled in the fit, is obtained from a generic MC sample and is
estimated by repeating the fit with and without this
background~(0.013, 0.038). The modes  $\Lambda^{+}_c \rightarrow
\Lambda X e^+ \nu$,  where $X$ represents additional decay
products, have never been observed. The current upper limit is $
\frac{ \mathcal{B}(\Lambda^{+}_c \rightarrow \Lambda X e^+ \nu) }
{ \mathcal{B}(\Lambda^{+}_c \rightarrow \Lambda e^+ \nu) } <
0.15$, where $X \ne 0$, at 90\% confidence
level~\cite{lambdacStudy}. The uncertainty due to the presence of
these modes is estimated from a series of fits, each having an
additional background component with floating normalization to
represent a $\Lambda^{+}_c \rightarrow \Lambda X e^+ \nu$ mode.
The uncertainty due to the possible presence of $\Lambda^{+}_c
\rightarrow \Lambda X e^+ \nu$ is assigned  as the largest
deviation from the main result found in these fits~(0.022, 0.091).
The uncertainty associated with the $\Lambda_{c}^+$ fragmentation
function is estimated by varying this function (0.003, 0.002). The
uncertainty related to MC modeling of the reconstruction
efficiency of slow pions produced in $\Lambda$ decays is obtained
by varying this efficiency according to our understanding of the
CLEO detector~(0.004, 0.003). For approximately 10\% of the data,
all from runs near the end of the CLEO II.V data taking period,  a
discrepancy between data and MC was found for $t> 0.8$.  The
effect of the discrepancy was determined by binning the data in
$t$ and performing a series of fits with a variable range of $t$
excluded.  The size of the systematic uncertainty is conservatively
taken to be the largest difference between the results of these fits 
and the main result (0.004, 0.072). 
The simulation used to obtain the main result does not include 
final state radiation. The systematic uncertainty due to this is 
determined as the difference between the result of a fit where 
electroweak radiative corrections have been 
modeled~\cite{photos} and the main result~(0.005, 0.009).

Adding all sources of systematic uncertainty in quadrature, the
final result is  $R = -0.31 \pm 0.05{\rm (stat)} \pm 0.04{\rm
(syst)}$ and $M_{\rm pole} = (2.21 \pm 0.08{\rm (stat)} \pm
0.14{\rm (syst)})$~GeV/$c^2$, 
the latter value being consistent with vector dominance. 
We also find $R = -0.35 \pm 0.04{\rm (stat)} \pm
0.04{\rm (syst)}$ from a fit with $M_{\rm pole} = m_{D_s^{*}}$.
Using the values of $R$ and $M_{\rm pole}$ obtained in the
simultaneous fit and the KK model, the mean value of the decay
asymmetry parameter of $\Lambda_{c}^{+} \rightarrow \Lambda e^+
\nu_e$~\cite{alphaLambdac} averaged over charge conjugate states
is calculated to be $\alpha_{\Lambda_{c}} =-0.86 \pm {0.03}{\rm
(stat)} \pm 0.02{\rm (syst)}$, for $\langle q^2 \rangle =
0.67$~(GeV/$c^2$)$^2$.

In the Standard Model, $CP$-violation is expected to be small in
semileptonic decays and absent in the decay $\Lambda_{c}^{+}
\rightarrow \Lambda e^+ \nu_e$. 
If $CP$ is conserved, the following relation is satisfied
\[
\frac{d\Gamma (\Lambda_{c}^{+} \rightarrow \Lambda e^+ \nu_e)}
{dq^{2}d\cos{\theta_{\Lambda}}d\cos{\theta_W}d\chi} =
\frac{d\Gamma (\overline{\Lambda}_{c}^{-} \rightarrow 
\overline{\Lambda} e^- \overline{\nu}_e)}
{dq^{2}d\cos{\theta_{\Lambda}}d\cos{\theta_W}d(-\chi)}. 
\]
Following~\cite{acp} and by extension, a $CP$-violating asymmetry
of the $\Lambda^+_c$ is defined as $\mathcal{A}_{\Lambda_{c}} =
\frac {(\alpha_{\Lambda_{c}} + \alpha_{\overline{\Lambda}_{c}})}
{(\alpha_{\Lambda_{c}} - \alpha_{\overline{\Lambda}_{c}})}$.
In the KK model, the angular distributions for $\Lambda_c^+$ and 
$\overline{\Lambda}_c^-$ are governed by $R$ and $M_{\rm pole}$. 
Using the values of $R$ and $M_{\rm pole}$ obtained 
in a simultaneous fit to each charge conjugate state separately
and the KK model, we calculate $\alpha_{\Lambda_{c}}
\alpha_{\Lambda} = -0.561 \pm 0.026{\rm (stat)}$ and
$\alpha_{\overline{\Lambda}_{c}} \alpha_{\overline{\Lambda}} =
-0.544 \pm 0.024{\rm (stat)} $.
From the following expression for the CP asymmetry:
$ \frac{\alpha_{\Lambda_{c}}\alpha_{\Lambda}  -
\alpha_{\overline{\Lambda}_{c}} \alpha_{\overline{\Lambda}}}
{\alpha_{\Lambda_{c}}\alpha_{\Lambda}  +
\alpha_{\overline{\Lambda}_{c}}\alpha_{\overline{\Lambda}}}
= \mathcal{A}_{\Lambda_{c}} + \mathcal{A}_{\Lambda}$, 
which is valid to first order in $\mathcal{A}_{\Lambda}$
and $\mathcal{A}_{\Lambda_{c}}$,
we obtain  $\mathcal{A}_{\Lambda_{c}} = 0.00 \pm 0.03{\rm (stat)}
\pm 0.01{\rm (syst)} \pm 0.02$,  where in
the systematic uncertainty the correlations among the systematic
uncertainties for the charge conjugate states are taken into
account and the third error is from the uncertainty in 
$\mathcal{A}_{\Lambda}$.


In conclusion, using a four-dimensional maximum likelihood fit, 
the angular distributions of $\Lambda_{c}^{+} \rightarrow \Lambda
e^+ \nu_e$ have been studied. The form factor ratio $R = f_2 /
f_1$ and $M_{\rm pole}$ are found to be $-0.31 \pm 0.05{\rm
(stat)} \pm 0.04{\rm (syst)}$ and $(2.21 \pm 0.08{\rm (stat)} \pm
0.14{\rm (syst)})$~GeV/$c^2$, respectively. This is the most
precise measurement of $R$, and it demonstrates that $f_2$ is 
non zero with a combined statistical and systematic 
significance exceeding~4$\sigma$. 
This is also the first measurement of $M_{\rm pole}$ in a charm  
baryon  semileptonic decay. 
Our measurement is consistent with vector dominance.
These results correspond to  $\alpha_{\Lambda_{c}} =-0.86
\pm {0.03}{\rm (stat)} \pm 0.02{\rm (syst)}$, for $\langle q^2
\rangle = 0.67$~(GeV/$c^2$)$^2$. Comparing the angular
distributions for $\Lambda_c^+$ and $\overline{\Lambda}_c^-$, 
no evidence for $CP$-violation is found: $\mathcal{A}_{\Lambda_{c}} =
0.00 \pm 0.03{\rm (stat)} \pm 0.01{\rm (syst)} \pm 0.02$, 
where the third error is from  the uncertainty in the world average 
of the $CP$-violating parameter, $\mathcal{A}_{\Lambda}$, 
for $\Lambda \rightarrow p \pi^-$.


We gratefully acknowledge the effort of the CESR staff 
in providing us with excellent luminosity and running conditions.
This work was supported by  the National Science Foundation,
the U.S. Department of Energy, the Research Corporation,
and the Texas Advanced Research Program.


\begin{thebibliography}{99}
\bibitem{PDG2004}
Particle Data Group, S. Eidelman {\em et al.,} Phys. Lett. B {\bf
593}, 1  (2004).


\bibitem{LambdaInHQET}
N. Isgur and M.B. Wise, Phys. Lett. B {\bf 232}, 113 (1989); {\bf 237}, 527 (1990);
E. Eichten and B. Hill, Phys. Lett. B {\bf 234}, 511 (1990);
H. Georgi, Phys. Lett. B {\bf 240}, 447 (1990).

\bibitem{heavyToLight}
T. Mannel, W. Roberts, and Z. Ryzak, Nucl. Phys. {\bf B355}, 38 (1991);
Phys. Lett. B {\bf 255}, 593 (1991);
F. Hussain, J.G. K\"{o}rner, M. Kr\"{a}mer, and G. Thompson, Z. Physics C {\bf 51}, 321 (1991);
A.F. Falk and M. Neubert, Phys. Rev D {\bf 47}, 2982 (1993);
H. Georgi, B. Grinstein and M.B. Wise, Phys. Lett. B {\bf 252}, 456 (1990);
N. Isgur and M.B. Wise, Nucl. Phys. {\bf B348}, 276 (1991);
H. Georgi, Nucl. Phys. {\bf B348}, 293 (1991).


\bibitem{KKModel} J.G. K\"{o}rner and M. Kr\"{a}mer, Phys. Lett. B {\bf 275},
495 (1992).


\bibitem{noteOnPlus}


The sign of the interference term (the term containing
$\cos{\chi}$) in Ref.~\cite{KKModel} has been corrected in
equation~\ref{rateTheory} with the approval of the authors of
Ref.~\cite{KKModel} (private communication).


\bibitem {cleo2} Y. Kubota {\it et al.,} Nucl. Instrum. Methods
Phys. Res. Sect. A {\bf 320}, 255 (1992).



\bibitem{cleo25} T. Hill, Nucl. Instrum. Methods
Phys. Res. Sect. A {\bf 418}, 32 (1998).


\bibitem{ffstudy}
CLEO Collaboration, G.L. Crawford  {\it et al.}, Phys. Rev. Lett.
{\bf 75}, 624 (1995).

\bibitem{lambdacStudy}  CLEO Collaboration, T. Bergfeld {\it et al.}, Phys. Lett. B {\bf 323},
219  (1994).


\bibitem{FWMoments}
    G.C.Fox and S.Wolfram {\em et al.}, Phys. Rev. Lett. {\bf 41} 1581 (1978).



\bibitem{cascadeCStudy}  CLEO Collaboration,
J. Alexander {\em et al.}, Phys. Rev. Lett. {\bf 74}, 3113 (1995).

\bibitem{fitMethod}  D. M. Schmidt, R.J. Morrison and M.S. Witherell, Nucl. Instrum.
Methods Phys. Res., Sect. A {\bf 328}, 547 (1993).

\bibitem{geant} R. Brun {\em et al.},  GEANT 3.15,
CERN Report No. DD/EE/84-1, 1987.

\bibitem{photos} E. Barberio and Z. Was, Comput. Phys. Commun. {\bf 79},
291 (1994).
D. Atwood and W. Marciano, Phys. Rev. D {\bf 41}, 1736 (1990)

\bibitem{alphaLambdac}
The decay asymmetry parameter $\alpha_{\Lambda_{c}}$ is defined
as~\cite{KKModel} $ \alpha_{\Lambda_{c}} = \frac { |H_{1/2~1}|^2 -
|H_{-1/2~-1}|^2 + |H_{1/2~0}|^2 - |H_{-1/2~0}|^2}
                          { |H_{1/2~1}|^2 + |H_{-1/2~-1}|^2 + 
|H_{1/2~0}|^2 + |H_{-1/2~0}|^2}. $

\bibitem{acp}
F. Donoghue, and S. Pakvasa, Phys. Rev. Lett. {\bf 55},
162 (1985).
A $CP$-violating parameter, $\mathcal{A}_{\Lambda}$, 
for $\Lambda \rightarrow p \pi^-$ is defined as  $\mathcal{A}_{\Lambda} 
\equiv \frac {(\alpha_{\Lambda} + \alpha_{\overline{\Lambda}})}
{(\alpha_{\Lambda} - \alpha_{\overline{\Lambda}})}$.
$\mathcal{A}_{\Lambda}$ is measured to be 
$\mathcal{A}_{\Lambda} =
0.012 \pm 0.021$~\cite{PDG2004}.

\end{thebibliography}
\end{document}